\documentclass[twocolumn]{aastex61}
\usepackage{apjfonts}

\received{}
\revised{}
\accepted{}
\submitjournal{ApJ}

\shorttitle{High Forbidden-to-resonance Line Ratio of \ion{O}{7} Discovered in the Cygnus Loop}
\shortauthors{Uchida et al.}

\begin{document}

\title{High Forbidden-to-resonance Line Ratio of  \ion{O}{7}  Discovered in the Cygnus Loop}

\author{H. Uchida}
\correspondingauthor{Hiroyuki Uchida}
\email{uchida@cr.scphys.kyoto-u.ac.jp}
\affil{Department of Physics, Graduate School of Science, Kyoto University, Kitashirakawa Oiwake-cho, Sakyo-ku, Kyoto 606-8502, Japan}

\author{S. Katsuda}
\affil{Graduate School of Science and Engineering, Saitama University, 255 Shimo-Ohkubo, Sakura, Saitama 338-8570, Japan}

\author{H. Tsunemi}
\affiliation{Department of Earth and Space Science, Osaka University, 1-1 Machikaneyama-cho, Toyonaka, Osaka 560-0043, Japan}

\author{K. Mori}
\affiliation{Department of Applied Physics and Electronic Engineering, Faculty of Engineering, University of Miyazaki, 1-1 Gakuen Kibanadai-Nishi, Miyazaki 889-2192, Japan}

\author{L. Gu}
\affiliation{RIKEN Nishina Center, 2-1 Hirosawa, Wako, Saitama 351-0198, Japan}
\affiliation{SRON Netherlands Institute for Space Research, Sorbonnelaan 2, 3584 CA Utrecht, the Netherlands}

\author{R. S. Cumbee}
\affiliation{Department of Physics and Astronomy and the Center for Simulational Physics, University of Georgia, Athens, GA 30602, USA}

\author{R. Petre}
\affiliation{X-ray Astrophysics Laboratory, NASA Goddard Space Flight Center, Greenbelt, MD 20771}

\author{T. Tanaka}
\affiliation{Department of Physics, Graduate School of Science, Kyoto University, Kitashirakawa Oiwake-cho, Sakyo-ku, Kyoto 606-8502, Japan}



\begin{abstract}
Charge exchange (CX) is an important process in shock physics since it indicates an interaction between downstream ions and ambient neutral hydrogen, suggesting a presence of a collisionless shock.
We present a high-resolution spectroscopy of an X-ray bright spot in a nearby supernova remnant (SNR), the Cygnus Loop, with the Reflection Grating Spectrometer (RGS) onboard \textit{XMM-Newton}.
The target is a compact knotty structure called ``southwestern knot (SW-K)'' located at the outer edge of the shell, where the blast wave is likely interacting with dense surrounding materials.
The RGS spectrum of the SW-K shows details of the line features below $\sim1$~keV, where we discover a high forbidden-to-resonance line ratio of \ion{O}{7} He$\alpha$.
The soft-band (10--35~\AA) spectrum is well explained by a thermal component with a CX X-ray emission obscured by neutral and ionized absorbers.
The presence of the CX X-ray emission will provide new insights into the shock physics of SNRs.
The high-resolution spectroscopy also reveals that the CNO, Ne and Fe abundances are truly lower than the solar values (0.2--0.4 solar) at the SW-K region .
Our result gives a clue to solving the previously known ``low-abundance problem'' reported from a number of evolved SNRs.
\end{abstract}

\keywords{ISM: abundances --- ISM: individual (Cygnus Loop) --- supernova remnants --- X-rays: ISM}



\section{Introduction} \label{sec:intro}

High resolution X-ray spectroscopy is a valuable tool for plasma diagnostics of astronomical objects.
One of those important aspects has been demonstrated by \textit{Hitomi} \cite[previously known as \textit{ASTRO-H};][]{Takahashi2016} observations of the Perseus cluster \citep{Hitomi2016, Hitomi2017}.
Other key targets for the high resolution spectroscopy are supernova remnants (SNRs) which are usually dominated by thermal (non-)equilibrium plasmas, especially when they are in an evolved stage.

The Cygnus Loop is a typical nearby \cite[540$^{+100}_{-80}$~pc; ][]{Blair2005} middle-aged \cite[$\sim14000$~yr; e.g.,][]{Levenson1998} SNR, which is considered to be a remnant of a core-collapse supernova explosion within a preexisting cavity \citep{Levenson1997}.
One of the interesting features in this context is that the Cygnus Loop is a Balmer-dominated SNR \citep{Raymond1983} where neutral atoms in interstellar material (ISM) are passing through a non-radiative shock front and directly interacting with hot postshock ions \citep{Ghavamian2001, Katsuda2016}.

Recent X-ray observations of the Cygnus Loop suggest a presence of charge exchange (CX) emission \citep{Katsuda2011CX} from part of the outer shell.
While several follow-up studies have been published so far, \citep[e.g.,][]{Leahy2013, Cumbee2014}, a study by \citet{Miyata2008} indicates that a resonance scattering is considerable for the shell of the Loop.  
Further direct observational evidence is required to close this issue.
In this paper, we report on a result of X-ray grating spectroscopy of the Cygnus Loop's outer edge where a CX emission has been suggested in a CCD spectrum \citep{Katsuda2011CX}. 
Throughout the paper, the errors are given at a 68\% confidence level unless otherwise noted.
While the distance to the Cygnus Loop is under debate \citep{Fesen2018}, we assume 540~pc \citep{Blair2005}  for our analysis unless otherwise noted.

\section{Observation and Data Reduction}
\citet{Katsuda2011CX} pointed out that evidence for the CX X-ray emission is found from outermost shells of the Cygnus Loop where a strong line emission is detected at $\sim0.7$~keV ($=17.7$~\AA), which is likely a complex of cascade lines of He-like oxygen (O He$\gamma$+$\delta$+$\epsilon$).
Among these regions, we selected a spot-like feature named ``southwestern knot (hereafter, SW-K)'' in the field of view (FOV) of the European Photon Imaging Camera \citep[EPIC;][]{Turner2001, Struder2001} onboard \textit{XMM-Newton} \citep{Jansen2001}.
We then targeted the SW-K with the Reflection Grating Spectrometer \citep[RGS;][]{denHerder2001} on 2012 May 9 during the Cycle 11 phase (Obs.ID 0693400101) so that we can perform a high-resolution X-ray spectroscopy: the angular size of the SW-K is roughly $\lesssim1$\arcmin, which is enough to resolve the multiplet lines of \ion{O}{7} and \ion{N}{6} with the RGS.

\begin{figure}[htb]
\begin{center}
\includegraphics[width=8cm]{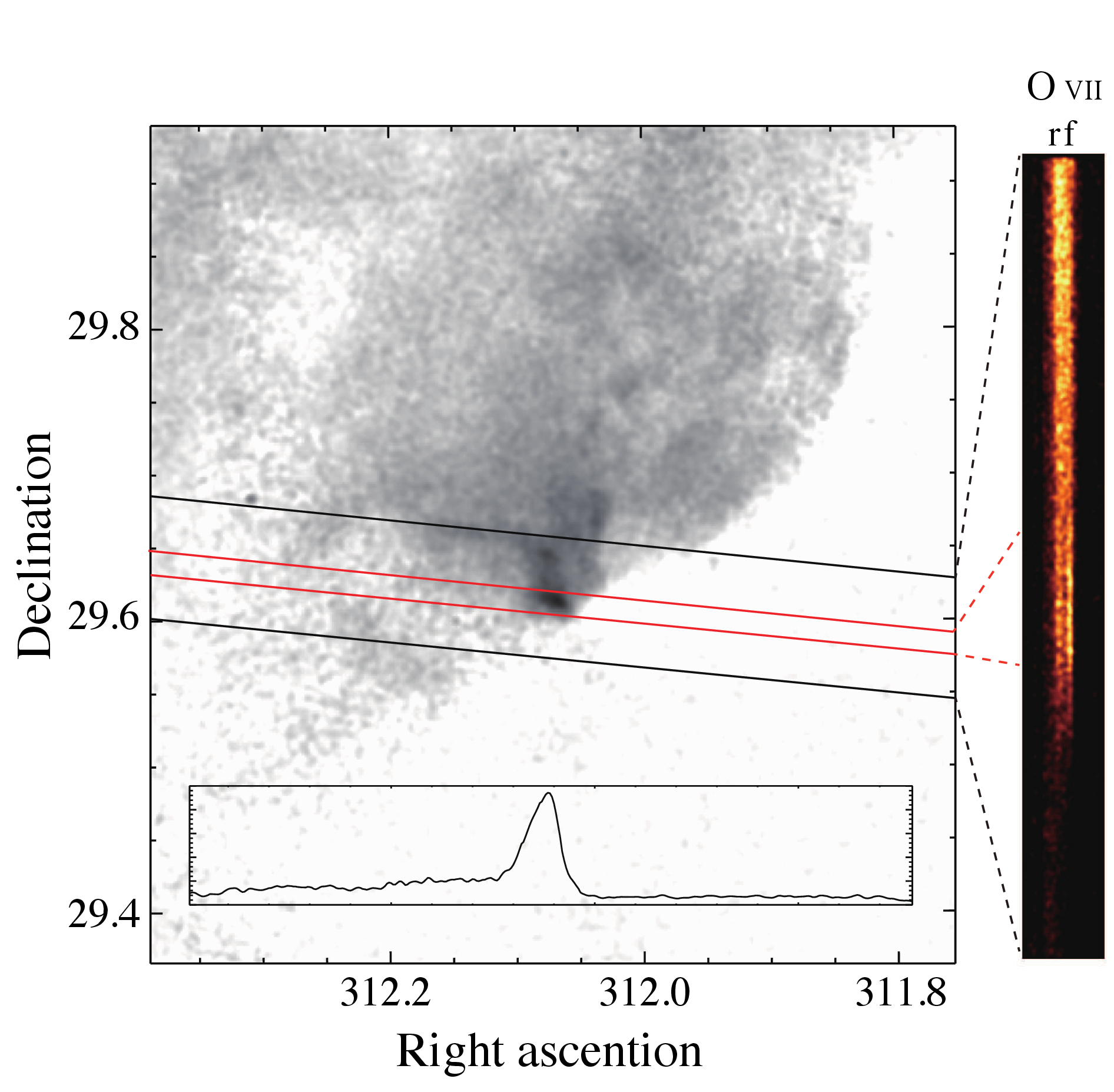}
\end{center}
\caption{ROSAT HRI image (0.1--2.4~keV) of the southwestern rim of the Cygnus Loop. 
The surface brightness is scaled logarithmically.
The cross-dispersion width of the RGS ($5\arcmin$) is in between the black lines.
The spectral extraction region for the SW-K is enclosed by the red lines.
The inset indicates the X-ray emission profile (in an arbitrary unit) along the RGS dispersion axis of the region from the right ascension of 311\arcdeg.8 to 312\arcdeg.4.
The right figure shows the RGS1 spatial image only displaying a dispersion angle corresponding to the \ion{O}{7} He$\alpha$ multiplet (20--24~\AA).
The notations, f and r represent the positions of the forbidden and resonance lines, respectively.
}\label{fig:rgs_image}
\end{figure}

The following analysis mainly focuses on the RGS data while the dataset obtained by the EPIC is supplementarily used to examine a broad band spectrum.
The RGS pipeline tool we used is \texttt{rgsproc} in the version 16.0.0 of the XMM Science Analysis Software.
We obtained total exposure times of 97.4~ks for both RGS1 and RGS2 after removing background flares according to the screening with the standard event selection criteria. 

As shown in the left panel of Figure~\ref{fig:rgs_image}, the dispersion direction of the RGS is 95.8\degr clockwise from north.
The FOV of RGS covers the outermost edge of the rim including the SW-K.
Using \texttt{rgsproc}, we generated an RGS1 spatial image of the \ion{O}{7} He$\alpha$ (20--24~\AA; right panel of Figure~\ref{fig:rgs_image}), which suggests an enhanced forbidden line of O$^{6+}$ in a region corresponding to the area of the SW-K.
We then extracted a spectrum by limiting the cross-dispersion angle between the red boundaries shown in Figure~\ref{fig:rgs_image}.

\section{Analysis}

\begin{figure*}[thb]
\plotone{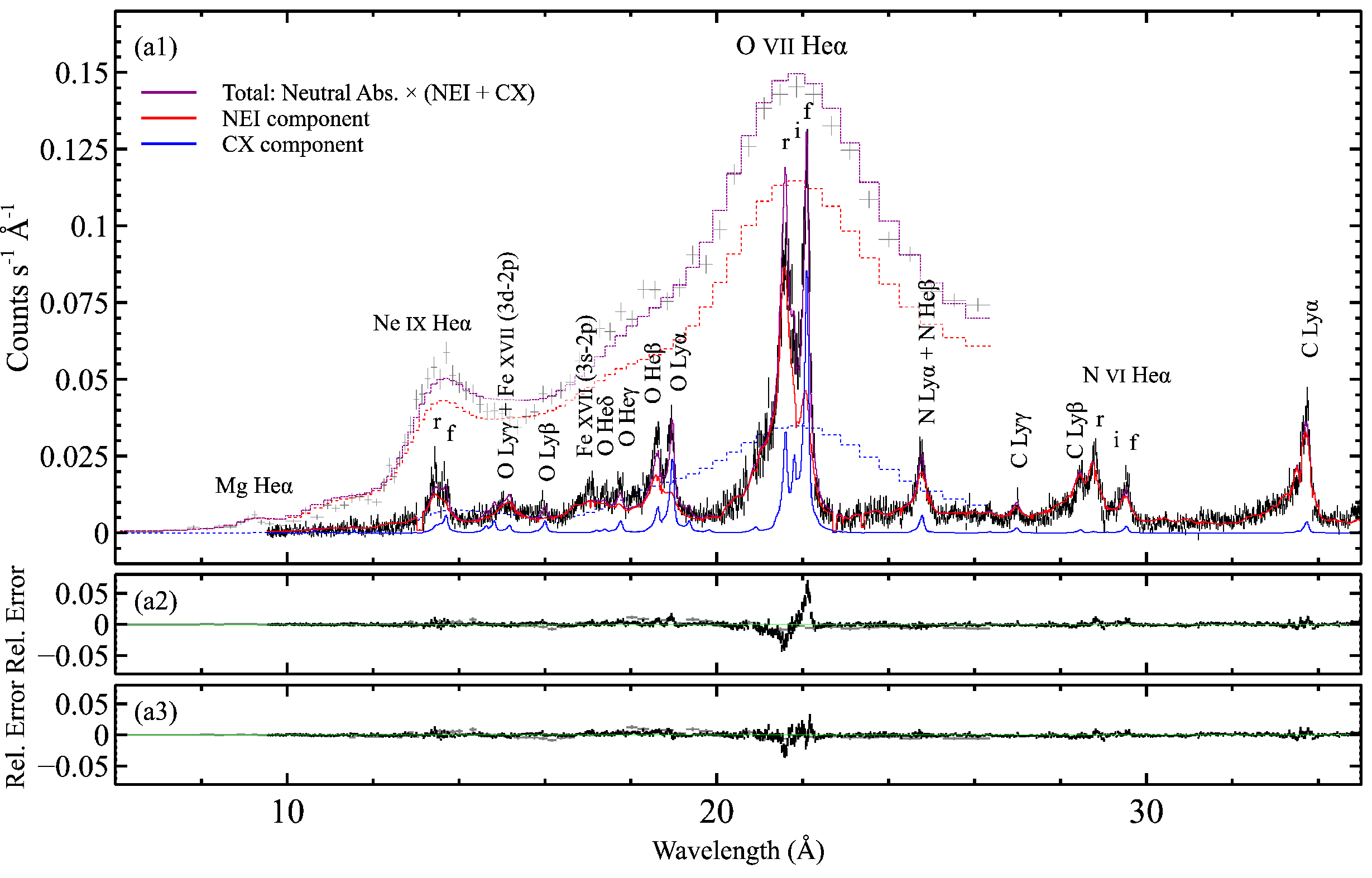}
\plotone{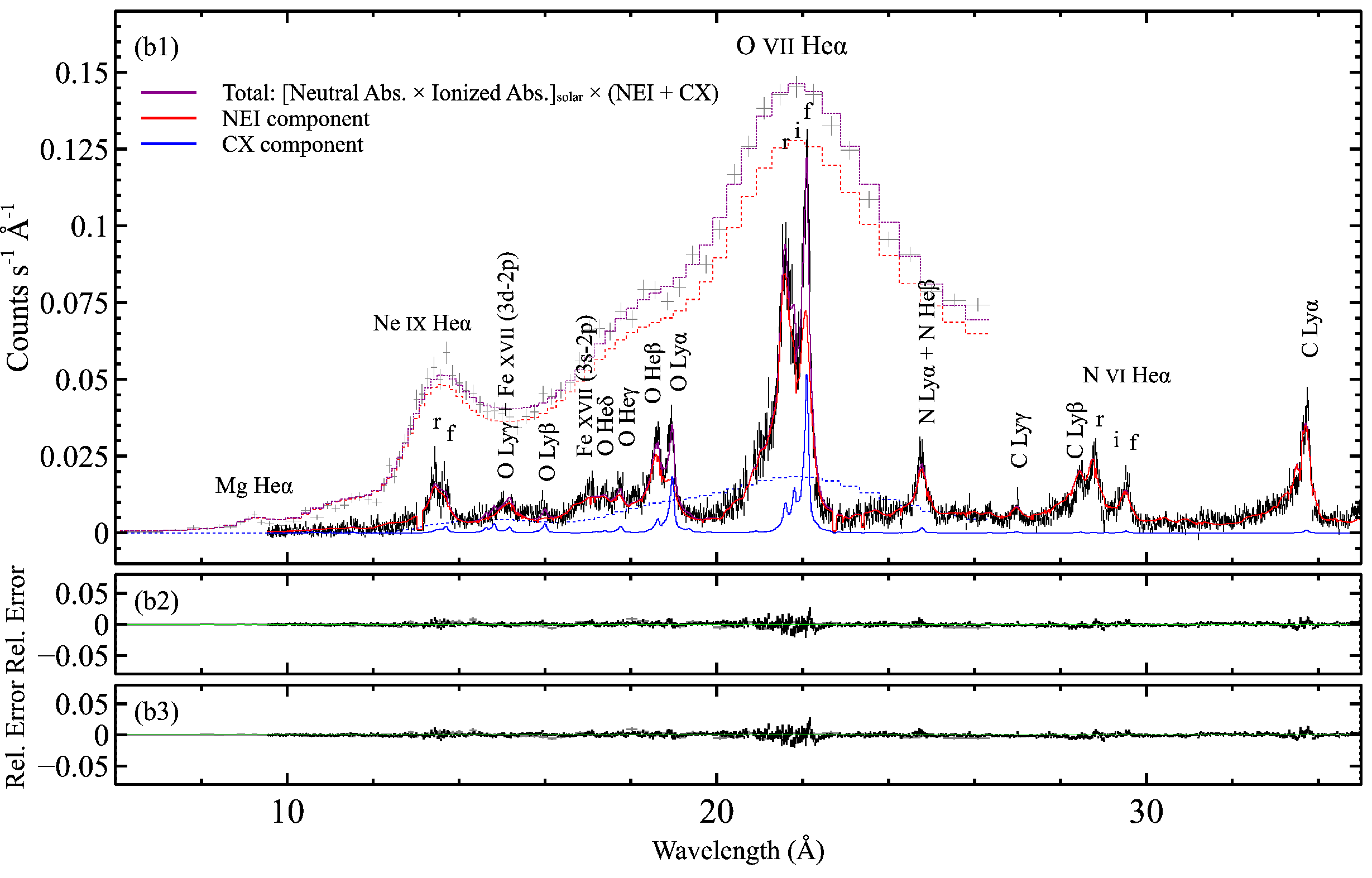}
\caption{RGS1$+$2 (black) and pn (gray) spectra of the SW-K. 
In panel~(a1), the solid and dotted lines represent the best-fit model of ``Neutral Abs. $\times$ ($\rm{NEI}+\rm{CX}$)'' for RGS and pn, respectively.
The residuals are shown in panel~(a3).
Residuals from the the best-fit model of the single NEI are shown in panel~(a2) for comparison.
Panel~(b1) shows the best-fit model of ``[Neutral Abs.$\times$Ionized Abs.]$_{\rm{solar}}$$\times$($\rm{NEI}+\rm{CX}$)''.
The residuals are shown in panel~(b2) whereas those for the model of ``Neutral Abs.$\times$[Ionized Abs.$\times$($\rm{NEI}+\rm{CX}$)]$_{\rm shell}$'' (see text and Table~\ref{tab:parameters}) are shown in panel~(b3).
}\label{fig:rgs_spectrum}
\end{figure*}

\begin{splitdeluxetable*}{lccccccBcccccccccccc}
\tabletypesize{\scriptsize}
\tablewidth{0pt} 
\tablenum{1}
\tablecaption{Best-fit parameters of the SW-K spectrum.\label{tab:parameters}}
\tablehead{
\colhead{Model Name} & $C$-$\rm{statistics/d.o.f.}$ &  \colhead{Component}& \colhead{$N_{\rm H}$} & \colhead{$kT_e$} & $n_e t$ &  $v_{\rm collision}$\tablenotemark{a} & rt\tablenotemark{b}& \multicolumn{7}{c}{Abundance} &  \colhead{Emission Measure\tablenotemark{c}}  \\
 \cline{9-15}  
  & \colhead{} & \colhead{} & \colhead{($10^{20}~\rm{cm^{-2}}$)} & \colhead{(keV)} & \colhead{($10^{10}~\rm{s~cm^{-1}}$)} &\colhead{($\rm{km~s^{-1}}$)} & & \colhead{C} & \colhead{N} & \colhead{O} & \colhead{Ne (=Mg)} & \colhead{Si} & \colhead{S (=Ar=Ca)} &  \colhead{Fe (=Ni)}  & \colhead{($10^{55}~\rm{cm^{-3}}$)} 
} 
\startdata 
Neutral Abs.$\times$NEI: (single NEI) &  $3941/1605$ & Neutral Abs.& 4.0 (fixed) & \nodata & \nodata  & \nodata &\nodata  &   \multicolumn{7}{c}{(fixed to the solar values)} &   \nodata    \\
  &&   NEI & \nodata & $0.213\pm0.004$ &  $5.2^{+0.7}_{-0.5}$ & \nodata  &\nodata & $0.22\pm0.02$  & $0.26\pm0.02$ & $0.16\pm0.01$ & $0.26\pm0.02$ &$0.7\pm0.1$&$0.33\pm0.04$ &  $0.30\pm0.02$&   $6.6^{+0.6}_{-0.5}$  \\
\hline
Neutral Abs.$\times$($\rm{NEI}+\rm{CX}$) & $3062/1603$ &  Neutral Abs. & 4.0 (fixed)  &  \nodata & \nodata & \nodata & \nodata & \multicolumn{7}{c}{(fixed to the solar values)} &   \nodata    \\
  &&  NEI& \nodata &  $0.224\pm0.003$ &  $3.6^{+0.3}_{-0.2}$ &  \nodata & \nodata  & $0.107^{+0.007}_{-0.006}$&  $0.15\pm0.01$ & $0.069\pm0.004$ & $0.126^{+0.008}_{-0.007}$  & $0.53^{+0.07}_{-0.06}$ & $0.18\pm0.02$ &  $0.26\pm0.02$ &  $8.7\pm0.5$ \\
&&  CX & \nodata & ($=\rm{NEI}$) &   \nodata & $<50$ & \nodata  &  \multicolumn{7}{c}{($=\rm{abundances~of~the~NEI~component}$)} & $73\pm1$ \\
\hline
[Neutral Abs.$\times$Ionized Abs.]$_{\rm{solar}}$$\times$NEI & $3009/1603$ &  Neutral Abs. &4.0 (fixed) & \nodata & \nodata &  \nodata  &  \nodata & \multicolumn{7}{c}{(fixed to the solar values)} &   \nodata    \\
&  &  Ionized Abs. & $0.18\pm0.01$ & $0.045^{+0.008}_{-0.007}$  &  \nodata & \nodata  & 1 (fixed) &  \multicolumn{7}{c}{(fixed to the solar values)} &      \nodata     \\
  &&  NEI& \nodata &  $0.226\pm0.004$ &  $3.9^{+0.4}_{-0.3}$ &  \nodata &\nodata &    $0.22^{+0.03}_{-0.01}$   & $0.33^{+0.04}_{-0.02}$ & $0.29^{+0.03}_{-0.01}$ &  $0.32^{+0.03}_{-0.02}$ &  $1.0^{+0.2}_{-0.1}$&  $0.41^{+0.06}_{-0.04}$   &  $0.42^{+0.04}_{-0.03}$ &   $4.7^{+0.3}_{-0.5}$ \\
\hline
 [Neutral Abs.$\times$Ionized Abs.]$_{\rm{solar}}$$\times$($\rm{NEI}+\rm{CX}$)& $2772/1601$ &  Neutral Abs. &4.0 (fixed) & \nodata & \nodata &  \nodata  &  \nodata & \multicolumn{7}{c}{(fixed to the solar values)} &   \nodata    \\
&  &  Ionized Abs. & $0.16^{+0.06}_{-0.04}$ &  $0.031^{+0.01}_{-0.002}$ &  \nodata & \nodata  & 1 (fixed) &  \multicolumn{7}{c}{(fixed to the solar values)} &      \nodata     \\
  &&  NEI& \nodata &  $0.240^{+0.005}_{-0.003}$ &  $2.5^{+0.2}_{-0.3}$ &  \nodata &\nodata &   $0.17\pm0.01$& $0.24\pm0.02$ & $0.16\pm0.01$ &  $0.20\pm0.01$ &  $0.7\pm0.1$&  $0.31\pm0.03$      &   $0.40^{+0.06}_{-0.02}$ & $5.5^{+0.4}_{-0.5}$\\
&&  CX & \nodata & ($=\rm{NEI}$) & \nodata & $<50$ & \nodata  &  \multicolumn{7}{c}{($=\rm{abundances~of~the~NEI~component}$)} &  $7\pm1$ \\
\hline
 Neutral Abs.$\times$[Ionized Abs.$\times$($\rm{NEI}+\rm{CX}$)]$_{\rm shell}$& $2794/1601$ &  Neutral Abs. & 4.0 (fixed)  & \nodata  & \nodata &  \nodata &  \nodata & \multicolumn{7}{c}{(fixed to the solar values)} &   \nodata    \\
&  &  Ionized Abs. & $0.7^{+0.1}_{-0.2}$ &  ($=\rm{NEI}$) &  \nodata & \nodata &  $0.45^{+0.05}_{-0.35}$ & \multicolumn{7}{c}{($=$ abundances of the NEI component)} &   \nodata    \\
  &&  NEI& \nodata &  $0.24^{+0.03}_{-0.05}$ &  $2.6^{+1.2}_{-0.3}$ &  \nodata  &  \nodata &  $0.20\pm0.02$ &  $0.26^{+0.18}_{-0.02}$ &   $0.18^{+0.02}_{-0.01}$ &  $0.24\pm0.02$ &  $0.8\pm0.1$ & $0.33^{+0.05}_{-0.04}$  & $0.40^{+0.58}_{-0.04}$  &  $5.0\pm0.5$ \\
&&  CX & \nodata & ($=\rm{NEI}$) & \nodata & $<226$ & \nodata  &  \multicolumn{7}{c}{($=\rm{abundances~of~the~NEI~component}$)} &   $1.3\pm0.3$ \\
\enddata
\tablenotetext{a}{Collision velocity between a hot plasma and cold neutral materials.}
\tablenotetext{b}{The ratio of ionization balance to electron temperature.}
\tablenotetext{c}{$\int n_{\rm e}n_{\rm H}$d$V$, where $n_{\rm e}$ and $n_{\rm H}$ are the electron and hydrogen densities and $V$ the volume of the source.}
\tablenotetext{d}{G-ratios are calculated from each unabsorbed model.}
\end{splitdeluxetable*}

\begin{table*}[!t]
\caption{G-ratios calculated from each unabsorbed model.\label{tab:g-ratio}}
\begin{center}
\begin{tabular}{lccccc}
\hline
\hline
 Model Name & &   & \multicolumn{3}{c}{Line Emission ($10^{40}$ photons~s$^{-1}$)} \\
  \cline{4-6}
&& G-$\rm{ratio}=(\rm{f}+\rm{i})/\rm{r}$& resonance & intercombination & forbidden  \\
\hline	   						    		      			  						
 Neutral Abs.$\times$NEI: (single NEI)  & NEI   & 0.774 & 6.953   &  1.107, 0.009 & 4.268\\ 
\hline	   						    		      			  						
 Neutral Abs.$\times$($\rm{NEI}+\rm{CX}$)  & NEI   & 0.749 & 5.007   &  0.769, 0.006 &2.974\\ 
  & CX  & 3.838 & 0.732 &  0.500, 0.004   & 2.306\\ 
\hline	   						    		      			  						
 [Neutral Abs.$\times$Ionized Abs.]$_{\rm{solar}}$$\times$NEI  & NEI   & 0.745 & 11.28   &  1.722, 0.013 &6.665\\ 
\hline	   						    		      			  						
 [Neutral Abs.$\times$Ionized Abs.]$_{\rm{solar}}$$\times$($\rm{NEI}+\rm{CX}$) & NEI   & 0.716 & 8.849   &  1.295, 0.010 &5.034\\ 
  & CX  & 3.838 & 0.444 &  0.303, 0.002   & 1.400\\ 
\hline	   						    		      			  						
Neutral Abs.$\times$[Ionized Abs.$\times$($\rm{NEI}+\rm{CX}$)]$_{\rm shell}$ & NEI   & 0.715 & 9.040   &  1.321, 0.010 &5.135\\ 
  & CX  & 3.732&  0.418   &  0.280, 0.002 &1.279\\ 
\hline
Data & & $1.79\pm0.09$ &  $5.3\pm0.2$   &  $1.1\pm0.2$ &$8.4\pm0.2$ \\
\hline	   						    		      			  							   						    		      			  						
\end{tabular}
\end{center}
\end{table*}

\subsection{High-resolution spectroscopy}
Hereafter, we apply version 3.03.0 of SPEX software \citep{Kaastra1996} in which we use the maximum likelihood \textit{C}-statistic for the following spectral analysis \citep{Cash1979, Kaastra2017}. 
To account for a spatial broadening of the RGS spectrum, we use a SPEX component, \texttt{Lpro}, which convolves models with a surface brightness profile estimated from the MOS1 image of the source.
Note that a scaling factor \textit{s}, which determines a broadening scales of the model, is decoupled for each component.

Figure~\ref{fig:rgs_spectrum}~(a1) shows the EPIC pn and a combined RGS1$+$2 spectra of SW-K, where prominent lines are found at $\sim13.5$~\AA \ (\ion{Ne}{9} He$\alpha$), 17--19~\AA \ (O He$\beta, \gamma, ...$ and Ly$\alpha$ with several L-shell lines of Fe), $\sim22$~\AA \ (\ion{O}{7} He$\alpha$), $\sim25$~\AA \ (N Ly$\alpha$ and He$\beta$), $\sim29$~\AA \ (\ion{N}{6} He$\alpha$) and $\sim34$~\AA \ (C Ly$\alpha$). 
Soft X-ray emission from the shell of the Cygnus Loop generally originates from shock-heated ISM as reported by several previous studies \citep{Tsunemi2007, Uchida2009LOS, Katsuda2011CX}.
We therefore tried a non-equilibrium ionization (NEI) model \citep{Kaastra1993} absorbed by neutral gas with cosmic abundances \citep{Morrison1983} for the spectra.
The absorption column density $N_{\rm H}$ was fixed to $4.0\times10^{20}~\rm{cm^{-2}}$ following previous measurements by \citet{Tsunemi2007}.
The electron temperature $kT_e$, ionization parameter $n_e t$ and emission measure of the NEI component were left free.
Abundances of C, N, O, Ne ($=\rm{Mg}$), and Fe ($=\rm{Ni}$) were also allowed to vary.
We also thawed Si, S ($=\rm{Ar}=\rm{Ca}$) abundances, although the L-shell emission from these elements is relatively negligible in the energy band of interest.
The others were fixed at the solar values.

\begin{figure}[t]
\begin{center}
\includegraphics[width=7.5cm]{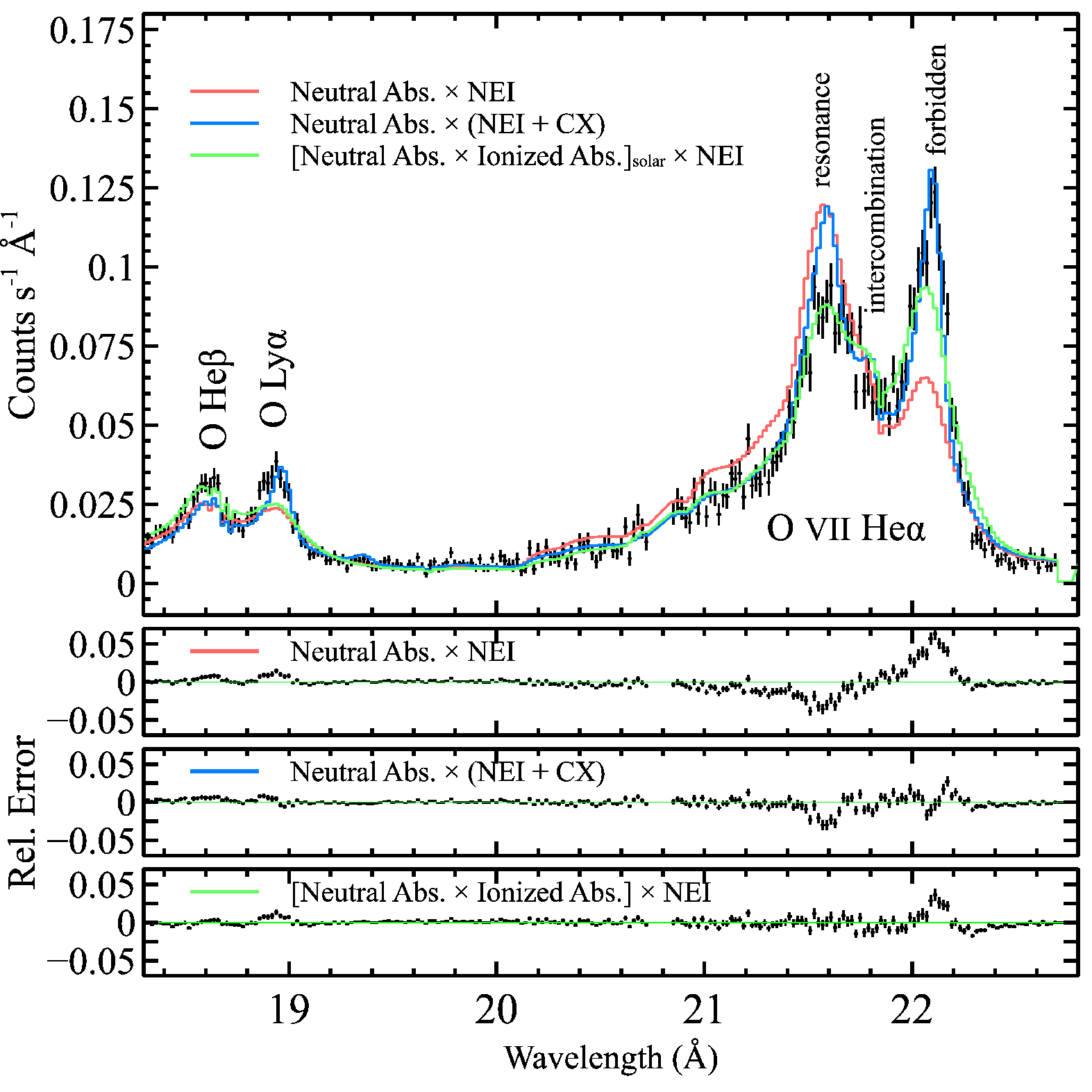}
\end{center}
\caption{Same RGS spectrum as figure~\ref{fig:rgs_spectrum}, but expanded to show \ion{O}{7} He$\beta$, \ion{O}{8} Ly$\alpha$ and \ion{O}{7} He$\alpha$.
The light red, blue, and green curves are the best-fit single-NEI, ``Neutral Abs. $\times$ ($\rm{NEI}+\rm{CX}$)'', and ``[Neutral Abs.$\times$Ionized Abs.]$_{\rm{solar}}$$\times$NEI'' models, respectively. 
Residuals are shown in the lower panels.}\label{fig:rgs_O_spectrum}
\end{figure}

While the best-fit parameters for the single-NEI model (Table~\ref{tab:parameters}) almost agree with those estimated from previous CCD observations of other shell regions of the Cygnus Loop \citep{Leahy2004, Tsunemi2007, Katsuda2008, Leahy2013}, we found a significant discrepancy between the model and data especially at the \ion{O}{7} He$\alpha$ line as shown in figures~\ref{fig:rgs_spectrum}~(a2) and \ref{fig:rgs_O_spectrum} with $C$-statistics of 3941 for 1605 degrees of freedom (d.o.f.).
We also display G-ratios obtained from the data and model in table~\ref{tab:g-ratio}.
Note that neither a collisional ionization equilibrium (CIE) nor a recombining plasma model can reproduce the broadband spectral feature including the intensity ratio of the \ion{O}{7} series.
We also confirmed that no improvement was achieved by using any combination of thermal plasmas (e.g., a two-component NEI with different $kT_e$, etc.).

To explain the observed \ion{O}{7} forbidden-to-resonance ratio, we then took into account a contribution of CX emission which may enhance triplet line intensity by effect of cascades following the CX transfer \citep{Katsuda2011CX, Cumbee2014, Roberts2015}.
A similar spectral feature is found from an RGS spectrum of Puppis~A by \citet{Katsuda2012}, who claimed that the anomalous line ratios can be explained by CX X-ray emission.
We thus added a CX model presented by \citet{Gu2016} to the NEI component, coupling all the abundances and $kT_{\rm e}$ to those of NEI, and varied the collision velocity $v_{\rm collision}$ which corresponds to a shock velocity in our case. 
Note that a better fit was obtained by assuming a multiple collision case in which an ion continuously undergoes CX until it becomes neutral.
The best-fit model and residuals are displayed in Figure~\ref{fig:rgs_spectrum}~(a1) and (a3), respectively. 
The parameters are summarized in Table~\ref{tab:parameters}.
Although the fit is improved with $C$-${\rm statistics/d.o.f.}=3062/1603$, the intensity ratios of \ion{O}{7} singlet/triplet cannot be fully reproduced.
As shown in figure~\ref{fig:rgs_O_spectrum}, the \ion{O}{7} He$\alpha$ and O Ly$\alpha$ are fairly reproduced by the blend of the NEI and CX components, whereas the \ion{O}{7} resonance line intensity is significantly overestimated.
We therefore conclude that the model ``Neutral Abs. $\times$ ($\rm{NEI}+\rm{CX}$)''  is insufficient to describe the spectrum of the SW-K.

Another possibility to account for the \ion{O}{7} line ratio is an ionized absorption proposed by \citet{Gu2016CX}. 
Based on the \textit{Suzaku} and \textit{XMM-Newton} analysis of the North Polar Spur spectra, they indicated that a foreground ionized absorber may explain the high forbidden-to-resonance ratio of \ion{O}{7}.
We thus applied a CIE absorption model, \texttt{hot}, which is available in SPEX \citep[cf.][]{dePlaa2004, Steenbrugge2005} to the best-fit $\rm{NEI}$ model.
The free parameters for the ionized absorption are the hydrogen column density and the electron temperature.
All the abundances were fixed at the solar values (hereafter, ``[Neutral Abs.$\times$Ionized Abs.]$_{\rm{solar}}$$\times$NEI'' model).
This model, however, again does not satisfy the whole line intensities of the \ion{O}{7} He$\alpha$ and O Ly$\alpha$ ($C$-${\rm statistics/d.o.f.}=3009/1603$).
As shown in figure~\ref{fig:rgs_O_spectrum}, the \ion{O}{7} resonance line is well fitted with the model, whereas the intensities of the forbidden and O Ly$\alpha$ lines are significantly underestimated.

We then employed the $\rm{NEI}+\rm{CX}$ model with the ionized absorption to mitigate the discrepancy between the ratios of the \ion{O}{7} and \ion{O}{8} line intensities.
This model led to a significant improvement in $C$-${\rm statistics/d.o.f.}=2772/1601$ as shown in figure~\ref{fig:rgs_spectrum}~(b1) and (b2). 
All the best-fit elemental abundances (table~\ref{tab:parameters}) are significantly lower than the solar values as reported in many previous studies \citep[e.g.,][]{Miyata1994, Leahy2004}: this point will be discussed in more detail in the following section.
The best-fit values of $kT_{\rm e}=0.240^{+0.005}_{-0.003}$~keV and $n_e t=(2.5^{+0.2}_{-0.3})\times10^{10}~\rm{s~cm^{-1}}$ are well within typical values for the shell component of this remnant \citep{Uchida2009Ejecta}.
We conclude that both the RGS and CCD spectra of the SW-K are reasonably fitted with this hybrid  model (namely, ``[Neutral Abs.$\times$Ionized Abs.]$_{\rm{solar}}$$\times$($\rm{NEI}+\rm{CX}$)'').

\subsection{Monochromatic Image}
The best-fit result explained above requires a different scaling factor \textit{s} for each component; NEI ($\sim1$) and CX ($\sim0.2$).
The result implies that the emitting region of the \ion{O}{7}  forbidden line is more compact than that of the resonance line.
In order to study a spatial distribution of \ion{O}{7} resonance and forbidden lines, we thus constructed a monochromatic image for each line in the same manner as described in \citet{Heyden2003}.

\begin{figure}[t]
\includegraphics[width=8.5cm]{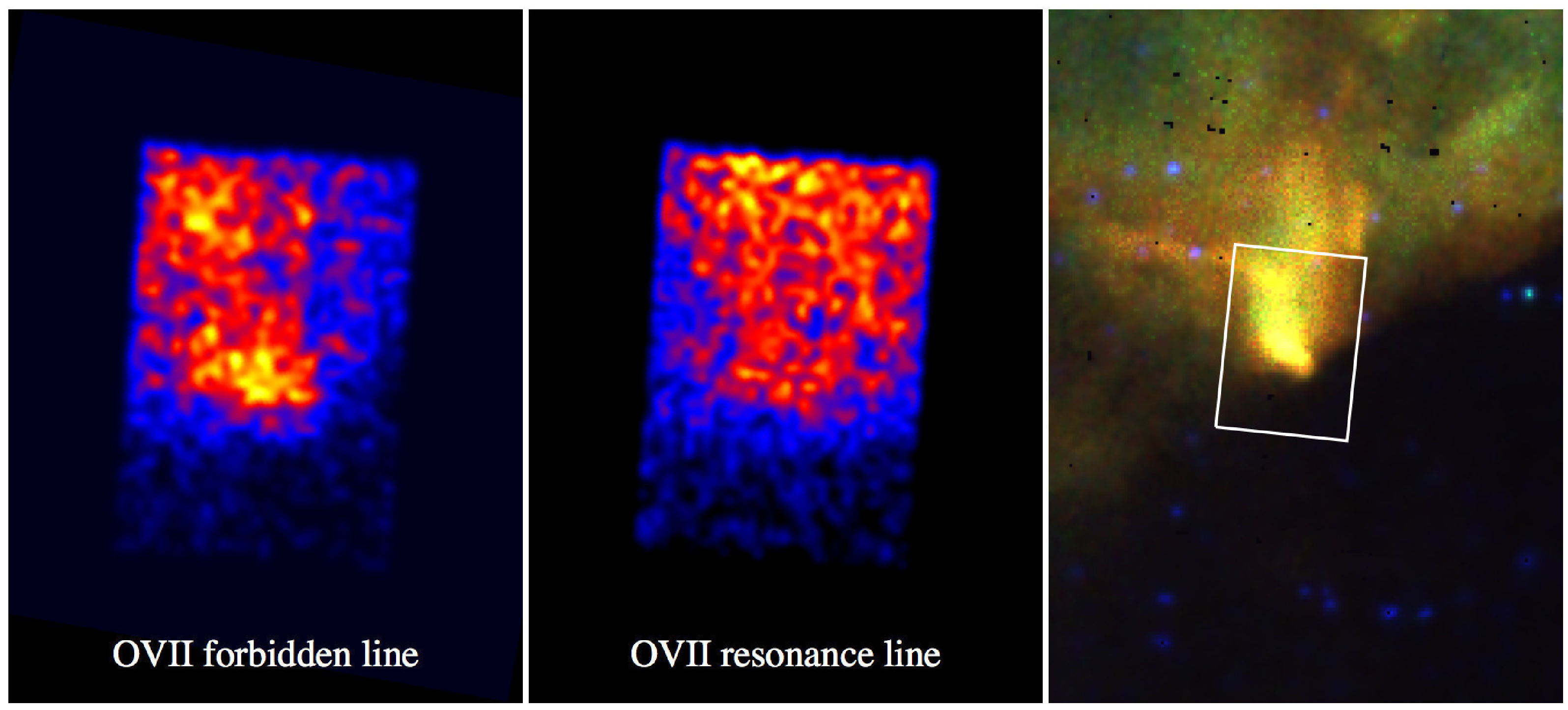}
\caption{\textit{Left}: RGS monochromatic image for the \ion{O}{7} forbidden line (22.1~\AA). \textit{Middle}: Same as the middle panel but for the \ion{O}{7} resonance line (21.6~\AA). \textit{Right}: True color image of SW-K obtained with the EPIC. Red, green, and blue represent 0.4--0.52~keV, 0.52--1.07~keV, and 1.07--0.3~keV, respectively. The white rectangular corresponds to the emitting regions of the left two images.
}\label{fig:rgs_mono}
\end{figure}

From figure~\ref{fig:rgs_mono}, the \ion{O}{7} forbidden line (22.1~\AA) image resolves two bright spots which correspond to SW-K and a northern small knot (see figure~\ref{fig:rgs_image}) whereas the \ion{O}{7} resonance line (21.6~\AA) image is more extended at the SW-K position.
We note that there are a CCD gap and the faint intercombination line emission on the right-hand side of the forbidden and resonance line images, respectively.
Nevertheless, figure~\ref{fig:rgs_mono} indicates that the emitting region of the \ion{O}{7} forbidden line is more compact than that of the resonance line.
The result suggests that the \ion{O}{7} forbidden line is enhanced at the SW-K \textit{or} the \ion{O}{7} resonance line is somehow weakened, as discussed in the following section.

\section{Discussion}

\subsection{Ionized Absorption}
As shown in the previous section, we found a satisfactory fit to the RGS spectra of the SW-K with the model,
\begin{equation}
 [\rm{Neutral~Abs.}\times\rm{Ionized~Abs.}]_{\rm{solar}}\times(\rm{NEI}+\rm{CX}).
\end{equation}
The absorbing column density of the ionized material $N_{\rm H~(hot)}$ is $\sim0.18\times10^{20}~\rm{cm^{-2}}$, while that of the neutral ISM $N_{\rm H~(neutral)}$ is $4.0\times10^{20}~\rm{cm^{-2}}$ (see Table~\ref{tab:parameters}).
The result suggests the presence of a cold ($<0.1~\rm{keV}$) ionized absorber that obscures the SW-K in the line of sight.
One of the promising candidates is the local hot bubble \citep[LHB;][]{Snowden1990} since the temperature of the absorber is close to that of the LHB \citep[$0.097\pm0.013$~keV;][]{Liu2017}.
However, assuming a line-of-sight scale of $<100~\rm{pc}$ and electron density of $\sim0.01~\rm{cm^{-3}}$ \citep{Smith2007}, we obtained its column density $N_{\rm H~(LHB)}<3\times10^{18}~\rm{cm^{-2}}$ for the LHB as an upper limit, which is almost negligible compared with the best-fit column density of $N_{\rm H~(hot)}$. 
The Galactic halo is also less plausible since its temperature \citep[$0.26\pm0.06$~keV;][]{Nakashima2018} is nearly ten times higher than that of our result.
We also rule out the possibility of absorption by the Cygnus superbubble \citep{Cash1980} because the Cygnus Loop is located in outermost and at the nearest position of the bubble \citep{Uyaniker2001}.

While we cannot rule out a possibility that unknown ionized gas exists along the line of sight between Earth and the SW-K region, here we consider another possibility that the ionized absorber is localized around the SW-K.
In this case, it is more reasonable to assume that the temperature and abundances of the absorber are coupled with those of the NEI/CX components.
Here we describe this model as,
\begin{equation}
 \rm{Neutral~Abs.}\times[\rm{Ionized~Abs.}\times(\rm{NEI}+\rm{CX})]_{\rm{shell}}, 
\end{equation}
where we varied the ratio of ionization balance to electron temperature of the ionized absorber to imitate the non-equilibrium ionization state.
While it gives a slightly worse fit than the previous one (Table~\ref{tab:parameters}), the \ion{O}{7} forbidden-to-resonance ratio is fairly explained even by this model.
The residuals are shown in Figure~\ref{fig:rgs_spectrum}~(b3).
The result suggests that a self absorption, or a resonance  scattering in some cases, is effective in the SW-K plasma,  although the above model is not equivalent to the definition of the resonance scattering.
We note that this possibility has been already pointed out by \citet{Miyata2008}.
The monochromatic images (figure~\ref{fig:rgs_mono}) hint that the \ion{O}{7} resonance line emission is obscured at the dense regions such as the SW-K and the northern small spot, which may be supporting evidence for this scenario.

Based on a far-ultraviolet observation of the remnant with the Hopkins Ultraviolet Telescope, \citet{Danforth2001} showed a considerable presence of the resonance scattering across the southeastern structure ``XA'', which is a bright spot of X-ray emission possibly enhanced by a shock-cloud interaction \citep{Hester1986, Zhou2010}.
It is remarkable that the SW-K has a similar structure of diameter $\lesssim1~\rm{pc}$ to XA and that both are indented on the edge of the Loop.
\citet{Danforth2001} indicated that a cloud of several-parsec scale in our line of sight is required to make such indented X-ray structure.
These results indirectly confirm the contribution of the self-absorption including the resonance scattering in the SW-K.

Here, we estimate a line-of-sight length of the absorber from the best-fit parameters of the model ``$\rm{Neutral~Abs.}\times[\rm{Ionized~Abs.}\times(\rm{NEI}+\rm{CX})]_{\rm{shell}}$''.
Since the model does not include the scattering effect, the estimation below gives an upper limit in the case that the resonance scattering is non-negligible.
Assuming a size of the SW-K is $1\arcmin\times2\arcmin\sim0.16~\rm{pc}\times0.32~\rm{pc}$ and a plasma depth of $\sim0.32~\rm{pc}$, we obtain a hydrogen density $n\rm{_H} \sim 5.5$$f^{-1/2}~\rm{cm^{-3}}$ from the best-fit parameter of the Emission Measure, where $f$ is a filling factor and we corrected for the spatial-integration factor of $\sim3.1$ by comparing the fluxes of RGS and EPIC spectra \citep[see also][]{Katsuda2012}.
If the ionized absorber is localized around the SW-K, its line-of-sight length is roughly estimated as $N_{\rm H~(hot)}/n\rm{_H}\sim4.2$$f^{1/2}~\rm{pc}$.
Given that the Loop is 25~pc in diameter and $f=1$, the line-of-sight length of $4.2~\rm{pc}$ is expected at $\sim0.4$~arcmin inside of the edge, which is consistent with the location of the SW-K.
The result supports that the ionized absorption is attributed to a low-temperature shell of the Cygnus Loop including the SW-K itself.

\subsection{Charge Exchange Emission}\label{sec:cx}
As described in the previous section (figure~\ref{fig:rgs_O_spectrum} and table~\ref{tab:parameters}), the observed \ion{O}{7} forbidden-to-resonance ratio cannot be reproduced by only considering the ionized absorption without the CX component.
We thus confirm that the SW-K spectrum requires the CX component.
The possible evidence for the CX X-ray emission is first pointed out by \citet{Katsuda2011CX} from X-ray observations of the Cygnus Loop shell with \textit{Suzaku}.
We note that the southwestern rim including the SW-K is one of the regions where \citet{Katsuda2011CX} have proposed a potential contribution of the CX X-ray emission.
The SW-K is located at the distorted shell near the ``blowout'' region \citep{Aschenbach1999, Uchida2008} and dominated by complex non-radiative (Balmer-dominated) filaments \citep{Roberts2015}.
Similar situations have been reported in other regions of the Cygnus Loop; the XA region \citep{Danforth2001}, a southwestern filament close to the SW-K \citep{Patnaude2005} and a southeastern rim ``P26'' \citep{Kosugi2010}.
Since a concentration of \ion{H}{1} gas is seen along the western edge of the Cygnus Loop \citep{Leahy2002}, it is reasonable to guess that the SW-K is a product of a recent interaction between the shock-heated SNR plasma and surrounding neutral materials.

We estimate a shock velocity of the SW-K from the best-fit parameter of $kT_e$ to be $\sim300~\rm{km~s^{-1}}$, which is a typical value in non-radiative shocks of the Loop \citep{Ghavamian2001}.
On the other hand, the obtained upper limit of the collision velocities shown in Table~\ref{tab:parameters} are much lower than the typical value but rather close to the decelerated shock velocity of $\sim170~\rm{km~s^{-1}}$ measured for XA by \citet{Danforth2001}.
Our result implies a deceleration of the shock around the SW-K.
A similar picture is proposed by \citet{Hester1994}, in which the shock velocity in the northeastern rim has decreased from  $\sim400~\rm{km~s^{-1}}$ to  $\sim180~\rm{km~s^{-1}}$ due to a collision with the surrounding dense cavity material. 
Based on an analogy of the northeastern rim and XA, we consider that the SW-K is also a product of such shock-cloud interaction.
The detected CX can occur in a complex region around the SW-K where the shock is quickly decelerated by the cavity material.
The monochromatic images shown in figure~\ref{fig:rgs_mono} may support this scenario as an alternative interpretation rather than the self absorption since the CX emission (i.e., \ion{O}{7} forbidden line emission) is enhanced around dense shock-cloud interaction regions such as the SW-K and the northern small spot (figure~\ref{fig:rgs_image}).

A previous study of the Cygnus Loop shows a strong resonance line of \ion{O}{7} with no forbidden line emission from a high-resolution spectrum obtained with the Focal Plane Crystal Spectrometer (FPCS)  on \textit{Einstein} \citep{Vedder1986}.
Since the aperture position of their observation is $\sim30\arcmin$ inside the northern edge of the Loop with FOV of $3\arcmin\times30\arcmin$, the X-ray emission is a superposition of metal-rich ejecta and projected swept-up ISM components.
Their result supports our speculation that CX interactions become effective only where the shock-heated ISM is interacting with surrounding neutral gas at the very edge of the shell.

\subsection{Low Abundance of Elements}
It has been widely known that abundances of the post-shock gas are substantially lower than the solar values in evolved SNRs: Puppis~A \citep{Hwang2008, Katsuda2010}, RCW~86 \citep{Vink1997}, G156.2$+$5.7 \citep{Uchida2012} and Vela SNR \citep{Miceli2008}, other than the Cygnus Loop.
As summarized by \citet{Long2014}, this ``low abundance problem'' might be attributed to some mechanisms such as a pollution from circumstellar material of the pre-SN star or heavy elements tied up in dust grains.
Another possibility they claimed is that non-standard plasma processes such as the CX emission can affect the inferred abundances because they are not taken into account in current standard SNR analyses. 
From Table~\ref{tab:parameters}, the best-fit model considering the CX X-ray emission results in a lower abundance (0.2--0.4 solar) for C, N, O, Ne and Fe.
We therefore conclude that the ISM abundance  in the vicinity of the Cygnus Loop is significantly depleted below the solar values.

As mentioned in section~\ref{sec:cx}, \citet{Katsuda2011CX}  showed possible evidence for the CX X-ray emission from the \textit{Suzaku} XIS spectra.
Their argument is mainly based on an excess emission at $\sim0.7~\rm{keV}$ found from CCD spectra that is possibly due to the complex of O He$\gamma+\delta+\epsilon$.
They claimed that the excess causes an apparent abundance enhancement if an X-ray spectrum is fitted with a thin-thermal model without the CX component.
We actually confirmed that a fit of the EPIC spectrum with an absorbed single NEI model gives an overestimation of the abundances ($\rm{N}=1.6\pm0.2$; $\rm{O}=0.73^{+0.08}_{-0.07}$).
However, we found no prominent excess at $\sim0.7~\rm{keV}$ (17.7~\AA) from the RGS spectra.
At least in the SW-K, the contribution of the O He$\gamma+\delta+\epsilon$ emission from the CX is not considerable.
Our result is consistent with a conclusion made by \citet{Cumbee2014}, who pointed out that CX emission models are insufficient to fully account for the 0.7 keV X-ray flux of the Cygnus Loop's shell.
To investigate the origin of the 0.7-keV excess, it is essential to perform high-resolution spectroscopy for regions where the emission is strongest, such as the outermost northeastern rim \citep{Katsuda2011CX}.

\section{Conclusions}
We observed a knotty structure located at the southern outermost edge of the Cygnus Loop, namely the SW-K, with the RGS onboard \textit{XMM-Newton}, which is the first high resolution spectroscopy of the rim-brightened region of this remnant.
The RGS resolves a soft-band X-ray spectrum that contains K-shell emission of C, N, O and Ne, where the forbidden-to-resonance line ratio of  \ion{O}{7}  is significantly higher than that expected from a standard thermal plasma emission.
From the high-resolution spectroscopy of the SW-K, we obtained three important new results as follows:

\begin{enumerate}
\item The CX contribution to the emission of the SW-K is non-negligible. 
The RGS monochromatic image indicates that the \ion{O}{7} forbidden line emission is enhanced around dense shock-cloud interaction regions.
If this is the case, our result provides evidence that post-shock ions are interacting with surrounding neutral gas, which will be important in the context of magnetic-field amplification related to collisionless plasma instabilities behind the shock \citep[e.g.,][]{Ohira2009}. 
\item Ionized absorption is required to account for the soft-band spectrum, which implies a presence of the hot Galactic ISM such as the Galactic halo. 
The most plausible explanation is, however, a local self-absorption \citep[or, in some cases, the resonance scattering;][]{Miyata2008} caused by the shell of the Cygnus Loop.
While its origin is still unclear, it can be easily tested through a future observation of an inner region of the Loop with a high resolution spectroscopy.
\item The ISM abundance of the SW-K is significantly lower than the solar value.
While the so-called ``low abundance problem'' of the evolved SNRs is widely known, a possibility of an underestimation of the line-to-continuum ratio has not been ruled out based on analysis of CCD data \citep{Long2014}.
The RGS spectrum clearly indicates that the low abundance problem is real, at least in the SW-K region of the Cygnus Loop.
\end{enumerate}

We note that it is expected to detect the CX X-ray emission from other outermost shells of the Loop as suggested by \citet{Katsuda2011CX} and \citet{Roberts2015}.
Since most of the candidate regions except the SW-K are too diffuse to resolve the multiplet lines with conventional grating spectrometers, an X-ray microcalorimeter \citep{Kelley2016} will be an excellent instrument for our purpose.
We have performed a feasibility study of this topic in the \textit{ASTRO-H} white paper \citep{Long2014}.
To confirm the presence of the CX X-ray emission in SNRs, it is essential to study a spatial variation of the CX contribution from outermost edge to inner shells, which will become one of major research topics for the near future missions such as the X-Ray Imaging and Spectroscopy Mission \citep[\textit{XRISM};][]{Tashiro2018} and the Advanced Telescope for High ENergy Astrophysics \citep[\textit{Athena};][]{Nandra2013}.

\acknowledgments
This work is supported by JSPS KAKENHI Grand Numbers 26800102 (H.U.) and 16K17673, 17H02864 (S.K.), 16H03983 (K.M.).
This work was partly supported by Leading Initiative for Excellent Young Researchers, MEXT, Japan.



\end{document}